\documentclass[10pt,rmp,notitlepage,final,preprint]{revtex4}%
\usepackage{amsfonts}
\usepackage{amsmath}
\usepackage{amssymb}
\usepackage{graphicx}
\usepackage{fancyhdr}%
\providecommand{\U}[1]{\protect\rule{.1in}{.1in}}

\begin{document}
\title{}
\maketitle

\begin{center}
{\Large Disorder, Path Integrals and Localization}

Gregg M. Gallatin

Applied Math Solutions, LLC

Newtown, CT

gregg@appliedmathsolutions.com

\bigskip
\end{center}

\begin{quotation}
{\small Anderson localization is derived directly from the path integral
representation of quantum mechanics in the presence of a random potential
energy function. The probability distribution of the potential energy is taken
to be a Gaussian in function space with a given autocorrelation function.
Averaging the path integral itself we find that the localization length, in
one-dimension, is given by }$\left(  E_{\xi}/\sigma\right)  \left(
KE_{cl}/\sigma\right)  \xi$ {\small where }$E_{\xi}$ {\small is the
"correlation energy", }$KE_{cl}$ {\small the average classical kinetic energy,
}$\sigma$ {\small the root-mean-square variation of the potential energy and
}$\xi$ {\small the autocorrelation length. Averaging the square of the path
integral shows explicitly that closed loops in the path when traversed
\ forward and backward in time lead to exponential decay, and hence
localization. We also show how, using Schwinger proper time, the path integral
result can be directly related to the Greens function commonly used to study
localization.}

\bigskip
\end{quotation}

\section{Introduction}

The fact that the wave function in a random or disordered potential is
localized in one and two dimensions and does not diffuse was predicted by
Anderson in the late 1950's $\left[  1\right]  $. Since then much work has
been done on "Anderson Localization", both theoretically and experimentally,
in a wide range of physical systems $\left[  2\right]  $. From the theoretical
perspective, localization is generally studied by considering the Greens
function $G\left(  \vec{x},\vec{x}^{\prime}\right)  =\langle\vec{x}|\left(
z-H\right)  ^{-1}|\vec{x}^{\prime}\rangle,$ or products of the Greens function
with itself, where $H$ is a Hamiltonian and $z$ is a complex number. In the
simplest case $H=\vec{p}^{2}/2m+V\left(  \vec{x}\right)  $ where the potential
$V\left(  \vec{x}\right)  $ is treated as disordered, i.e., as a random
function of position. Computing the average of $G\left(  \vec{x},\vec
{x}^{\prime}\right)  $ over some distribution of $V\left(  \vec{x}\right)  ,$
or powers of $G\left(  \vec{x},\vec{x}^{\prime}\right)  ,$ with $V\left(
\vec{x}\right)  $ in the denominator is rather complicated and as pointed out
in "Quantum Field Theory in a Nutshell" by A. Zee $\left[  3\right]  $ that
has led to two main approaches for studying localization using field theory
techniques, the replica approach pioneered by Parisi and others $\left[
4\right]  $ and the supersymmetry approach pioneered by Efetov $\left[
5\right]  $. But other approaches have also been used, see for example the
self-consistent approach developed by W\"{o}lfle and Vollhardt, Chapter 4 in
$\left[  2\right]  $.

Here we show how localization emerges directly from the path integral for
nonrelativistic quantum mechanics and also show how the path integral result
can be related to the above Greens function. Averaging the path integral over
a Gaussian probability distribution, in function space, for the potential
energy $V\left(  \vec{x}\right)  $ with a Hamiltonian of the form $H=\vec
{p}^{2}/2m+V\left(  \vec{x}\right)  $ is straightforward. This same approach
was used by Dashen in $\left[  6\right]  .$ There he considered the
propagation of light in a medium with a randomly varying index of refraction
in the paraxial approximation. His results therefore are equivalent to the 2D
Schrodinger equation with $\left(  x,y\right)  $ being position and the $z$
direction being time. Since he treats the randomness in the index of
refraction as spatially isotropic his analysis doesn't allow for the
randomness to be independent of $z$ which is required to get localization.
Hence, although localization is implicit in his analysis, he doesn't identify
it as such. But, since he does include real temporal variation in the index of
refraction he does find some very interesting "speckle" and "twinkle" effects.
Chakravarty and Schmid $\left[  7\right]  $ studied localization by
considering the path integral from a semiclassical point of view which
illustrated explicitly how closed loops in the path contribute to
localization. Here the path integral for a random potential is analyzed from
the fully quantum mechanical viewpoint. The average of the square of the path
integral shows explicitly the contribution of closed loops.

Using the Schwinger proper time representation of the inverse of a time
independent operator $\left[  8\right]  $ we can write, for $\operatorname{Im}%
\left[  z\right]  >0,$ \
\begin{align}
G\left(  \vec{x},\vec{x}^{\prime}\right)   &  =\frac{1}{i\hbar}\int%
_{0}^{\infty}dTe^{izT/\hbar}\left\langle \vec{x}\left\vert \exp\left[
-i\frac{HT}{\hbar}\right]  \right\vert \vec{x}^{\prime}\right\rangle
\nonumber\\
&  \equiv\frac{1}{i\hbar}\int_{0}^{\infty}dTe^{izT/\hbar}K\left(  \vec
{x},T,\vec{x}^{\prime},0\right)
\end{align}
The inclusion of $\hbar$ is not necessary but it does allow for the
interpretation of $T$ as actual time. This can be rewritten in terms of a path
integral $\left[  3,9\right]  $ as%
\begin{equation}
K\left(  \vec{x},T,\vec{x}^{\prime},0\right)  =\int_{\vec{x}^{\prime},0}%
^{\vec{x},T}\delta\vec{x}\left(  t\right)  \exp\left[  \frac{i}{\hbar}\int%
_{0}^{T}dt\left(  \frac{m}{2}\left(  \partial_{t}\vec{x}\left(  t\right)
\right)  ^{2}-V\left(  \vec{x}\left(  t\right)  \right)  \right)  \right]
\end{equation}
where $\int_{\vec{x}^{\prime},0}^{\vec{x},T}\delta\vec{x}\left(  t\right)  $
indicates integration over all paths $\vec{x}\left(  t\right)  $ starting at
$\vec{x}$ at $t=0$ and ending at $\vec{x}^{\prime}$ at $t=T.$ As is well known
$K\left(  \vec{x},T,\vec{x}^{\prime},0\right)  $ propagates a wave function,
$\psi\left(  \vec{x},t\right)  ,$ in time, i.e.,%

\begin{equation}
\psi\left(  \vec{x},T\right)  =\int d^{N}x^{\prime}K\left(  \vec{x},T,\vec
{x}^{\prime},0\right)  \psi\left(  \vec{x}^{\prime},0\right)
\end{equation}
where $N$ is the space dimension. Here we consider localization from the point
of view of the path integral itself. The point of (1) is to show how results
obtained using the path integral, which can easily be averaged over $V\left(
\vec{x}\right)  $ with a Gaussian probability distribution, can be directly
related to the more standard Greens function analysis of localization. That
relationship will not be explored in detail here.

Take $V\left(  \vec{x}\right)  $ to be a random function with a Gaussian
probability distribution $P\left[  V\left(  \vec{x}\right)  \right]  $ in
function space given by
\begin{equation}
P\left[  V\left(  \vec{x}\right)  \right]  =\frac{1}{Z_{0}}\exp\left[
-\frac{1}{2}\int d^{N}xd^{N}x^{\prime}V\left(  \vec{x}\right)  C^{-1}\left(
\vec{x},\vec{x}^{\prime}\right)  V\left(  \vec{x}^{\prime}\right)  \right]
\end{equation}
Here $N$ is the number of space dimensions and $Z_{0}$ is a normalization
factor which we won't need. We will generally use the notation $\int d\vec{x}$
instead of $\int d^{N}x.$

Expectation values with respect to $P\left[  V\left(  \vec{x}\right)  \right]
$ are defined by
\begin{equation}
\left\langle F\left(  V\left(  \vec{x}\right)  \right)  \right\rangle
_{P}=\int\delta V\left(  \vec{x}\right)  F\left[  V\left(  \vec{x}\right)
\right]  P\left[  V\left(  \vec{x}\right)  \right]
\end{equation}
where $\int\delta V\left(  \vec{x}\right)  $ indicates functional integration
over $V\left(  \vec{x}\right)  $. We find via standard procedures $\left[
3\right]  ,$
\begin{align}
\left\langle V\left(  \vec{x}\right)  \right\rangle _{P}  &  =0\nonumber\\
\left\langle V\left(  \vec{x}\right)  V\left(  \vec{x}^{\prime}\right)
\right\rangle _{P}  &  =C\left(  \vec{x},\vec{x}^{\prime}\right)
\end{align}
where $C\left(  \vec{x},\vec{x}^{\prime}\right)  $ is the inverse of
$C^{-1}\left(  \vec{x},\vec{x}^{\prime}\right)  $, i.e.,
\begin{equation}
\int d\vec{x}^{\prime\prime}C^{-1}\left(  \vec{x},\vec{x}^{\prime\prime
}\right)  C\left(  \vec{x}^{\prime\prime},\vec{x}^{\prime}\right)
=\delta\left(  \vec{x}-\vec{x}^{\prime}\right)
\end{equation}
Although not necessary, we will choose $C\left(  \vec{x},\vec{x}^{\prime
}\right)  $ to be isotropic and homogeneous,
\begin{equation}
C\left(  \vec{x},\vec{x}^{\prime}\right)  =C\left(  \left\vert \vec{x}-\vec
{x}^{\prime}\right\vert \right)
\end{equation}
and we will use the exponential form for the autocorrelation function%
\begin{equation}
\text{ }C\left(  x,\vec{x}^{\prime}\right)  =\sigma^{2}\exp\left[
-\frac{\left\vert \vec{x}-\vec{x}^{\prime}\right\vert }{\xi}\right]
\end{equation}
where $\sigma^{2}=\left\langle V\left(  \vec{x}\right)  ^{2}\right\rangle
_{P}$ is the mean square variation in the potential energy with $\xi$ the
autocorrelation length of $V\left(  \vec{x}\right)  $.

Again via thoroughly standard procedures $\left[  3\right]  $ we have
\begin{equation}
\left\langle \exp\left[  i\int d\vec{x}J\left(  \vec{x}\right)  V\left(
\vec{x}\right)  \right]  \right\rangle _{P}=\exp\left[  -\frac{1}{2}\int
d\vec{x}d\vec{x}J\left(  \vec{x}\right)  C\left(  \vec{x},\vec{x}^{\prime
}\right)  J\left(  \vec{x}^{\prime}\right)  \right]
\end{equation}
Note that this is properly normalized since
\begin{equation}
\left\langle 1\right\rangle _{P}=\left.  \exp\left[  -\frac{1}{2}\int d\vec
{x}d\vec{x}J\left(  \vec{x}\right)  C\left(  \vec{x},\vec{x}^{\prime}\right)
J\left(  \vec{x}^{\prime}\right)  \right]  \right\vert _{J\left(  \vec
{x}\right)  =0}=1
\end{equation}
so an explicit form for $Z_{0}$ is not needed.

Averaging the product of the wave function at one position with its conjugate
at another position, $\left\langle \sum_{i}\delta\left(  E-E_{i}\right)
\psi_{i}\left(  \vec{x}\right)  ^{\ast}\psi_{i}\left(  \vec{x}^{\prime
}\right)  \right\rangle _{P},$ shows how the wave function, at energy $E,$
correlates with itself as a function of $\left\vert \vec{x}-\vec{x}^{\prime
}\right\vert $. Here the $E_{i}$ are the energy eigenvalues and $\psi
_{i}\left(  \vec{x}\right)  $ the eigenfunctions, respectively, of $H.$ But,
as pointed out in $\left[  3\right]  ,$ since $\psi\left(  \vec{x}\right)  $
has a phase, this average potentially could yield zero. Instead it is
suggested in $\left[  3\right]  $ to average the square of the wave function
at two positions, i.e.,
\begin{equation}
\left\langle \sum_{i}\delta\left(  E-E_{i}\right)  \left\vert \psi_{i}\left(
\vec{x}\right)  \right\vert ^{2}\left\vert \psi_{i}\left(  \vec{x}^{\prime
}\right)  \right\vert ^{2}\right\rangle _{P}%
\end{equation}
Here we consider more generally the behavior of both the average of the
product of the wave function at different times and positions which leads to
averaging the path integral itself, and the behavior of the average of the
product of the square of the wave functions at two \ different times and
positions which leads to averaging the square of the path integral.

As pointed out above, the path integral evolves the wave function in time.
Consider a wave function initially localized on a scale much smaller than
$\xi,$ e.g., $\psi\left(  \vec{x},0\right)  =\exp\left[  -\frac{\vec{x}^{2}%
}{4w^{2}}\right]  /\left(  2\pi w^{2}\right)  ^{1/4}$ with $w\ll\xi$ so that
$\left\vert \psi\left(  \vec{x},0\right)  \right\vert ^{2}\sim\delta\left(
\vec{x}\right)  $ on the scale of $\xi,$ and evaluate how it correlates with
itself at later times and positions. Using the chosen $\psi\left(  \vec
{x},0\right)  $, setting $\vec{x}_{e}=\vec{x}$ and $\vec{x}_{s}=\vec
{x}^{\prime}$ we have
\begin{align}
\left\langle \psi\left(  \vec{x},T\right)  \psi\left(  0,0\right)  ^{\ast
}\right\rangle _{P}  &  =\int d^{N}x^{\prime}\left\langle K\left(  \vec
{x},T,\vec{x}^{\prime},0\right)  \right\rangle _{P}\psi\left(  \vec{x}%
^{\prime},0\right)  \psi\left(  0,0\right)  ^{\ast}\nonumber\\
&  \sim\left\langle K\left(  \vec{x},T,0,0\right)  \right\rangle _{P}%
\end{align}
which is the average of the path integral itself. As shown below this does not
yield zero and in fact shows evidence of localization, at least when evaluated
for the classical trajectory. In line with evaluating the average of the
product of the square of the wave function at two positions, we also consider
\begin{align}
\left\langle \left\vert \psi\left(  \vec{x},T\right)  \right\vert
^{2}\left\vert \psi\left(  0,0\right)  \right\vert ^{2}\right\rangle _{P}  &
=\int d^{N}x^{\prime}d^{N}x^{\prime\prime}\left\langle K\left(  \vec{x}%
,T,\vec{x}^{\prime},0\right)  ^{\ast}K\left(  \vec{x},T,\vec{x}^{\prime\prime
},0\right)  \right\rangle _{P}\psi\left(  \vec{x}^{\prime},0\right)  ^{\ast
}\psi\left(  \vec{x}^{\prime\prime},0\right)  \left\vert \psi\left(
0,0\right)  \right\vert ^{2}\nonumber\\
&  \sim\left\langle \left\vert K\left(  \vec{x},T,0,0\right)  \right\vert
^{2}\right\rangle _{P}%
\end{align}
which, for the chosen $\psi\left(  \vec{x},0\right)  ,$ we see is
approximately the average of the square of the path integral.

\section{Averaging the Path Integral}

Letting $J\left(  \vec{x}\right)  =-\frac{1}{\hbar}\int_{0}^{T}dt\delta\left(
\vec{x}-\vec{x}\left(  t\right)  \right)  $ we have%
\begin{align}
\left\langle \exp\left[  -\frac{i}{\hbar}\int_{0}^{T}dtV\left(  \vec
{x}\right)  \right]  \right\rangle _{P}  &  =\left\langle \exp\left[  i\int
d^{N}xJ\left(  \vec{x}\right)  V\left(  \vec{x}\right)  \right]  \right\rangle
_{P}\nonumber\\
&  =\exp\left[  -\frac{1}{2\hbar^{2}}\int_{0}^{T}dtdt^{\prime}C\left(
\left\vert \vec{x}\left(  t\right)  -\vec{x}\left(  t^{\prime}\right)
\right\vert \right)  \right]
\end{align}
and so
\begin{equation}
\left\langle K\left(  \vec{x},T,\vec{x}^{\prime},0\right)  \right\rangle
_{P}=\int_{\vec{x}^{\prime},0}^{\vec{x},T}\delta\vec{x}\left(  t\right)
\exp\left[  \frac{i}{\hbar}\int_{0}^{T}dt\frac{m}{2}\left(  \partial_{t}%
\vec{x}\left(  t\right)  \right)  ^{2}-\frac{1}{2\hbar^{2}}\int_{0}%
^{T}dtdt^{\prime}C\left(  \left\vert \vec{x}\left(  t\right)  -\vec{x}\left(
t^{\prime}\right)  \right\vert \right)  \right]
\end{equation}
Thus averaging over $V\left(  \vec{x}\right)  $ has replaced the phase factor
$\exp\left[  -\frac{i}{\hbar}\int_{0}^{T}dtV\left(  \vec{x}\left(  t\right)
\right)  \right]  $ with the exponential factor. $\exp\left[  -\frac{1}%
{2\hbar^{2}}\int_{0}^{T}dtdt^{\prime}C\left(  \left\vert \vec{x}\left(
t\right)  -\vec{x}\left(  t^{\prime}\right)  \right\vert \right)  \right]  .$
After accounting for differences in notation this is the same as equation
(2.5) in Dashen $\left[  6\right]  $. He considers this result to be
exponentially small and "therefore not particularly interesting."\ We now show
that this result indicates the existence of localization. We work in one
dimension for simplicity.

Let $x^{\prime}=0$ and $x=L.$ The phase factor $\exp\left[  \frac{i}{\hbar
}\int_{0}^{T}dt\frac{m}{2}\left(  \partial_{t}^{2}x\left(  t\right)  \right)
^{2}\right]  $ is stationary when $x\left(  t\right)  $ is the classical
solution $x_{cl}\left(  t\right)  =\frac{L}{T}t$. Also the exponential factor
$\exp\left[  -\frac{1}{2\hbar^{2}}\int_{0}^{T}dtdt^{\prime}C\left(  \left\vert
x\left(  t\right)  -x\left(  t^{\prime}\right)  \right\vert \right)  \right]
$ is smaller the more times $x\left(  t\right)  $ intersects itself, i.e., the
more times for which $\left\vert x\left(  t\right)  -x\left(  t^{\prime
}\right)  \right\vert \ll\xi.$ Both these facts indicate that the dominant
contribution to the path integral comes from the straightest or classical path
$x_{cl}\left(  t\right)  =\frac{L}{T}t.$ Substituting this into exponential
factor and using the explicit form for $C\left(  \left\vert \vec{x}-\vec
{x}^{\prime}\right\vert \right)  $ we find%
\begin{align}
\left\langle K\left(  L,T,0,0\right)  \right\rangle _{P}  &  \simeq\sqrt
{\frac{m}{2\pi i\hbar T}}\exp\left[  i\frac{mL^{2}}{2\hbar T}-\frac{\sigma
^{2}\xi^{2}T^{2}}{\hbar^{2}L^{2}}\left(  \frac{\left\vert L\right\vert }{\xi
}+e^{-\left\vert L\right\vert /\xi}-1\right)  \right] \nonumber\\
&  \simeq\sqrt{\frac{m}{2\pi i\hbar T}}\exp\left[  i\frac{mL^{2}}{2\hbar
T}-\frac{\sigma^{2}\xi}{\hbar^{2}\left(  L/T\right)  ^{2}}\left\vert
L\right\vert \right]  \text{ for }L\gg\xi
\end{align}
where we have included the standard factor $\sqrt{\frac{m}{2\pi i\hbar T}}%
$.\ The exponential decay with increasing $\left\vert L\right\vert $ indicates
localization with the localization length $\ell$ proportional to the factor
$\hbar^{2}\left(  L/T\right)  ^{2}/\sigma^{2}\xi.$ Multiplying and dividing by
$\xi m/2$ and rearranging we have
\begin{equation}
\frac{\hbar^{2}\left(  L/T\right)  ^{2}}{\sigma^{2}\xi}=2\frac{\left(
\frac{\hbar^{2}}{m\xi^{2}}\right)  }{\sigma}\frac{\left(  \frac{m}{2}\left(
\frac{L}{T}\right)  ^{2}\right)  }{\sigma}\xi=2\frac{E_{\xi}}{\sigma}%
\frac{KE_{cl}}{\sigma}\xi\equiv2\ell
\end{equation}
where $E_{\xi}=\hbar^{2}/m\xi^{2}$ is the so called "correlation
energy"$~\left[  10\text{-}12\right]  $. It is proportional to $\left(
\lambda_{dB}/\xi\right)  ^{2}$ where $\lambda_{dB}$ is the de Broglie
wavelength. Hence $E_{\xi}$ is large when $\lambda_{dB}\gg\xi.$ In this regime
the wave function can spread over long distances via tunneling $\left[
\text{10-12}\right]  $ and leads to larger values of $\ell$. $KE_{cl}%
=m/2\left(  L/T\right)  ^{2}$ is the classical kinetic energy for the path
$x\left(  t\right)  =Lt/T$. Since $\left\langle V\left(  x\right)
\right\rangle _{P}=0$ we have the classical kinetic energy is equal to the
average total energy, $KE_{cl}=\left\langle E_{total}\right\rangle _{P}.$ The
Gaussian form for $C\left(  \left\vert \vec{x}-\vec{x}^{\prime}\right\vert
\right)  $ yields the same result for $L\gg\xi$ but without the factor of 2.

The justification for replacing $\left(  L/T\right)  ^{2}$ with $\left(
2/m\right)  KE_{cl}$ is because the path integral is a point to point
propagator, hence it includes all possible energies. Localization on the other
hand is usually studied at specific energy scales, e.g., the Fermi energy. We
can filter the path integral for specific average kinetic energies, $KE_{ave}$
by inserting a Fadeev-Popov type factor $\left[  3\right]  $
\begin{equation}
1=\int_{0}^{\infty}d\left(  KE_{ave}\right)  \delta\left(  KE_{ave}-\frac
{1}{T}\int_{0}^{T}dt\frac{m}{2}\left(  \partial_{t}x\left(  t\right)  \right)
^{2}\right)
\end{equation}
in the path integral. This is similar to the approach used by Chakravarty and
Schmid $\left[  7\right]  .$

\section{Averaging the $\left\vert \text{Path Integral}\right\vert ^{2}$}

Consider the average of the product of the square of the wave function at two
times, 0 and $T,$ and positions 0 and $\vec{L}$, which, as shown above, for
the chosen initial wave function reduces to\
\begin{equation}
\left\langle \left\vert \psi\left(  \vec{L},T\right)  \right\vert
^{2}\left\vert \psi\left(  0,0\right)  \right\vert ^{2}\right\rangle _{P}%
\sim\left\langle K\left(  \vec{L},T,0,0\right)  ^{\ast}K\left(  \vec
{L},T,0,0\right)  \right\rangle _{P}%
\end{equation}
Expressing one propagator as a path integral over $\vec{x}_{1}\left(
t\right)  $ and the other over $\vec{x}_{2}\left(  t\right)  $ where both
$\vec{x}_{1}\left(  t\right)  $ and $\vec{x}_{2}\left(  t\right)  $ go from
$\vec{x}_{s}=0$ to $\vec{x}_{e}=\vec{x}$, and letting $J\left(  \vec
{x}\right)  =-\frac{1}{\hbar}\int_{0}^{T}dt\left(  \delta\left(  \vec{x}%
-\vec{x}_{1}\left(  t\right)  \right)  -\delta\left(  \vec{x}-\vec{x}%
_{2}\left(  t\right)  \right)  \right)  $ we have
\begin{align}
&  \left\langle K\left(  \vec{L},T,0,0\right)  ^{\ast}K\left(  \vec
{L},T,0,0\right)  \right\rangle _{P}\nonumber\\
&  =\int_{0,0}^{\vec{L},T}\delta\vec{x}_{1}\left(  t\right)  \delta\vec{x}%
_{2}\left(  t\right)  \left\{  \exp\left[  \frac{i}{\hbar}\int_{0}^{T}%
dt\frac{m}{2}\left[  \left(  \partial_{t}\vec{x}_{2}\left(  t\right)  \right)
^{2}-\left(  \partial_{t}\vec{x}_{1}\left(  t\right)  \right)  ^{2}\right]
\right]  \right. \nonumber\\
&  \times\left.  \exp\left[  -\frac{1}{2\hbar^{2}}\int_{0}^{T}dtdt^{\prime
}\left(
\begin{array}
[c]{c}%
C\left(  \left\vert \vec{x}_{1}\left(  t^{\prime}\right)  -\vec{x}_{1}\left(
t\right)  \right\vert \right) \\
+C\left(  \left\vert \vec{x}_{2}\left(  t^{\prime}\right)  -\vec{x}_{2}\left(
t\right)  \right\vert \right) \\
-2C\left(  \left\vert \vec{x}_{2}\left(  t^{\prime}\right)  -\vec{x}%
_{1}\left(  t\right)  \right\vert \right)
\end{array}
\right)  \right]  \right\}
\end{align}
\ \ After accounting for differences in notation the is the same as equation
(2.9) in Dashen $\left[  6\right]  .$ Dashen notes that a dominant
contribution to the path integral comes from paths for which $\vec{x}%
_{1}\left(  t\right)  \simeq\vec{x}_{2}\left(  t\right)  ,$ and he uses that
in his evaluation$.$ But he does not note that paths with closed loops
traversed in opposite directions, equivalent to forward and backward in time
around the loop, also make a dominant contribution to the path integral. For
such paths $\left\vert \vec{x}_{2}\left(  t\right)  -\vec{x}_{1}\left(
t\right)  \right\vert $ can be arbitrarily large. Consider identical $\vec
{x}_{1}\left(  t\right)  $ and $\vec{x}_{2}\left(  t\right)  $ with one or
more closed loops. If both loops are traversed in the same direction this is
still simply $\vec{x}_{1}\left(  t\right)  =\vec{x}_{2}\left(  t\right)  .$
But for any of the loops traversed in the opposite direction then, with the
loop starting at $t=t_{a}$ and ending at $t=t_{b},$ we have $\vec{x}%
_{2}\left(  t^{\prime}\right)  =\vec{x}_{1}\left(  t_{a}+t_{b}-t^{\prime
}\right)  $ for $t_{a}\leq t^{\prime}\leq t_{b}$. \ Changing the integration
variable from $t^{\prime}$ to $\ t^{\prime\prime}=t_{a}+t_{b}-t^{\prime}$ for
$t_{a}\leq t^{\prime}\leq t_{b}$ gives the same result as $\vec{x}_{1}\left(
t\right)  =\vec{x}_{2}\left(  t\right)  $ in both exponential factors in the
path integral. \ Hence even though for oppositely traversed loops $\left\vert
\vec{x}_{2}\left(  t\right)  -\vec{x}_{1}\left(  t\right)  \right\vert $ can
be arbitrarily large, we still get a unity contribution to the path
integral$.$ The fact that forward and backward loops contribute equally to the
path integral is simply due to the fact that the accrued phase along the paths
in the forward and backward directions around each loop is the same and so the
two directions add coherently $\left[  2,5,7\right]  $. This fact is captured
in the very form of the second exponential.

It is not possible to evaluate the double path integral in an exact analytical
way. So we need to consider the character of the paths that make dominant contributions.

The first exponential is unity for any $\vec{x}_{2}\left(  t\right)
\not =\vec{x}_{1}\left(  t\right)  $ if the two paths have identical time
average kinetic energy$,$
\begin{equation}
\frac{m}{2}\int_{0}^{T}dt\left(  \partial_{t}\vec{x}_{2}\left(  t\right)
\right)  ^{2}=\frac{m}{2}\int_{0}^{T}dt\left(  \partial_{t}\vec{x}_{1}\left(
t\right)  \right)  ^{2}%
\end{equation}
Hence, from a stationary phase point of view, since $\left\langle V\left(
\vec{x}\right)  \right\rangle _{p}=0,$ the first exponential "filters out"
paths with very different average energies or, equivalently, very different
average de Broglie wavelengths.

To see how localization emerges, we must show that, in the second exponential
the first two terms, on average, outweigh the third term. First note that for
any paths whatsoever, the first two terms in the integrand are maximal for all
$t^{\prime}=t$ and so their integrals will scale with $T.$ In the third term
this can only happen if $\left\vert \vec{x}_{2}\left(  t^{\prime}\right)
-\vec{x}_{1}\left(  t\right)  \right\vert \ll\xi$ for all $t^{\prime}\simeq t$
or if both paths contain the same closed loops which again may be traversed in
either direction. But for very different paths, i.e., paths for which
$\left\vert \vec{x}_{2}\left(  t^{\prime}\right)  -\vec{x}_{1}\left(
t\right)  \right\vert \gg\xi$ \ for almost all $t^{\prime}$ and $t,$ then,
after accounting for closed loops, the third term will generally only be
nonzero for specific pairs or ranges of $t$ and $t^{\prime}$ where the paths
might cross or get close to one another, hence, on average, the first two
terms dominate the path integral and lead to localization.

Consider 1D. Both paths must cover the distance from 0 to $L$ in time $T.$ If
both paths remain between 0 and $L$ for $0\leq t\leq T,$ then the third term
and first two terms scale the same since there are multiple ranges of times
for which $x_{2}\left(  t^{\prime}\right)  =x_{1}\left(  t\right)  $ with
$t^{\prime}\not =t.$ But for paths where, say, $x_{2}\left(  t\right)  $
spends a significant portion of time at positions $x>L$ while $x_{1}\left(
t\right)  $ spends a significant portion of time at positions $x<L,$ or vice
versa, then, in this case the contribution from the third term will be much
less than that from the first two terms simply because there are now far fewer
ranges of time for which $x_{2}\left(  t^{\prime}\right)  =x_{1}\left(
t\right)  .$

Dimensionality plays a role in localization in the following way. Localization
occurs for any energy in one and two dimensions. In three dimensions there is
a "mobility edge" in terms of energy which separates extended from localized
wave functions $\left[  2,3,5,13,14\right]  .$ This is encapsulated in the
double path integral above as follows$.$ The path integral integrates over all
continuous random walks. But random walks in 1D and 2D are recurrent, i.e.,
every random walk in 1D and 2D will eventually cross itself somewhere,
creating closed loops. This was proven by P\'{o}lya in 1921 $\left[
15\right]  $. In 1D this is because any change in direction makes the path
recurrent and creates a closed loop with a range of times, $t$ and $t^{\prime
},$ for which $x_{2}\left(  t^{\prime}\right)  =x_{1}\left(  t\right)  $ with
$t^{\prime}\not =t$. In 2D recurrence occurs because the path itself forms a
boundary which, when crossed, again creates a closed loop. In this case
generally you have individual values of $t$ and $t^{\prime},$ for which
$\vec{x}_{2}\left(  t^{\prime}\right)  =\vec{x}_{1}\left(  t\right)  $ with
$t^{\prime}\not =t.$ Effectively, paths with closed loops are "dense" in path
space in 1D and 2D. The probability of a path being recurrent in 3D is less
than unity and paths with closed loops are, effectively, not "dense" in path
space. The P\'{o}lya result is the flip side of the Poincare-Bendixson theorem
$\left[  16,17\right]  $ which states that an autonomous dynamical system in
continuous time needs a phase space of at least 3 dimensions to have chaotic
trajectories, i.e., trajectories that never intersect themselves while
remaining in a finite volume of the phase space.

Finally note that if we let $V\left(  \vec{x}\right)  $ be random in time as
well as space, i.e., $V\left(  \vec{x}\right)  \rightarrow V\left(  \vec
{x},t\right)  ,$ then the autocorrelation function becomes $C\left(
\left\vert \vec{x}-\vec{x}^{\prime}\right\vert ,\left\vert t-t^{\prime
}\right\vert \right)  .$ Given a finite autocorrelation time $\tau,$ if
$T\ll\tau$ we have $C\left(  \left\vert \vec{x}-\vec{x}^{\prime}\right\vert
,\left\vert t-t^{\prime}\right\vert \right)  \simeq C\left(  \left\vert
\vec{x}-\vec{x}^{\prime}\right\vert ,0\right)  $ and we have the same results
as above, i.e., localization. But for $T\gg\tau$, peaks in $V\left(  \vec
{x},t\right)  $ evolve into valleys and vice versa which leads to diffusion.
From the point of view of the time integrals over the autocorrelation
function, the temporal variation of $V\left(  \vec{x},t\right)  $ destroys the
phase coherence of oppositely traversed closed loops for times $T\gg\tau.$
Effectively the relevant time scale for these integrals switches from $T$ to
$\tau,$ causing localization to degenerate into diffusion $\left[  2,7\right]
.$

\section{Acknowledgement}

The author would like to acknowledge helpful email exchanges with Alan Chodos
and Dieter Vollhardt.

\section{References}

\begin{enumerate}
\item Anderson, P. W., "Absence of Diffusion in Certain Random Lattices",
Phys. Rev. \textbf{109}, 1492 (1958).

\item Elihu Abrahams, Ed., 2010, \textit{50 Years of Anderson Localization},
World Scientific Publishing Co. Pte. Ltd., Singapore.

\item Anthony Zee, 2010, \textit{Quantum Field Theory in a Nutshell}, 2nd.
Ed., Princeton University Press, Princeton, NJ, USA.

\item Marc Mezard, Giorgio Parisi, Miquil Angel Virasoro, Eds., 1987,
\textit{Spin Glass Theory and Beyond}, World Scientific Publishing Co. Pte.,
Ltd. Singapore.

\item Konstatin Efetov, 1997, \textit{Supersymmetry in Disorder and Chaos},
Cambridge University Press, Cambridge, UK.

\item Dashen, R., "Path Integrals for Waves in Random Media", Journal of
Mathematical Physics, \textbf{20}, 894, (1979).

\item Chatravarty, S., Schmid, A., "Weak Localization: The quasiclassical
theory of electrons in a random potential", Physics Reports, \textbf{140},
Number 4, 193-236, (1986).

\item Schwinger, J., "On Gauge Invariance and Vacuum Polarization", Physical
Review, \textbf{82}, 664 (1951).

\item R. P. Feynman, A. R. Hibbs, 1965, \textit{Quantum Mechanics and Path
Integrals}, McGraw-Hill, Inc., New York, NY, USA.

\item Pasek, M., Orso, G., and Delande, D., "Anderson localization of
ultracold atoms: Where is the mobility edge?", Phys. Rev. Lett. 118.170403 (2017).

\item Kuhn, R. C., Miniatura, C., Delande, D., Sigwarth, O., and Muller, C. A.
"Localization of matter waves in two dimensional disordered optical
potentials", Phys. Rev. Lett. 95, 250403 (2005).

\item Kuhn, R. C., Sigwarth, O., Miniatura, C., Delande, D. and Muller, C. A.
"Coherent matter wave transport in speckle potentials", New J. Phys. 9, 161 (2007).

\item Gorkov, L. P., Larkin, A. I., Khmelnitskii, D. E., "Particle
Conductivity in a Two-Dimensional Random Potential", Pis'ma Zh. Eksp. Teor.
Fiz. \textbf{30}, 248 (1979), (Sov. Phys. JETP Lett. \textbf{30}, 228).

\item Anderson, P. W., Abrahams, E., Ramakrishnan, T. V., "Scaling Theory of
Localization: Absence of Quantum Diffusion in Two Dimensions", Phys. Rev.
Lett. \textbf{43}, 718 (1979).

\item P\'{o}lya, G. "Uber eine aufgabe betreffend die irrfahrt im
strassennetz.
\"{}
Math. Ann., \textbf{84},149--160 (1921).

\item Poincar\'{e}, H., "Sur les courbes d\'{e}finies par une \'{e}quation
diff\'{e}rentielle", Oeuvres, vol. 1, Paris, 1892.

\item Bendixson, I, "Sur les courbes d\'{e}finies par des \'{e}quations
diff\'{e}rentielles", Acta Mathematica \textbf{24}, 1--88 (1901).
\end{enumerate}

\end{document}